\pdfoutput=1
\documentclass[aps,prl,nofootinbib,preprintnumbers,amsmath,amssymb,latexsym,array,enumerate,letter,twocolumn]{revtex4}
\usepackage{mathtools}
\usepackage{epsfig}
\usepackage{graphicx}
\usepackage{color}
\usepackage{amssymb}
\usepackage{enumitem}
\usepackage{amsmath}
\usepackage{hyperref}
\usepackage{breakurl}

\makeatletter

\def\be{\begin{equation}}
\def\ee{\end{equation}}
\newcommand{\bea}{\begin{eqnarray}}
\newcommand{\eea}{\end{eqnarray}}

\begin{document}

\preprint{DESY-25-151}

\title{Did our Universe Tunnel out of the Wrong Higgs Vacuum?}

\author{Bibhushan Shakya}

\affiliation{ \vskip 2mm William H.\ Miller III Department of Physics \& Astronomy, Johns Hopkins University, 3400 N.\ Charles St., Baltimore, MD 21218, USA}

\affiliation{Deutsches Elektronen-Synchrotron DESY, Notkestr.\,85, 22607 Hamburg, Germany}

\begin{abstract}
This paper explores various aspects and implications of the initial configuration of the Standard Model (SM) Higgs field at the beginning of our Universe. It is well known that the SM Higgs field features a deeper, more stable minimum at large field values. While it is possible that our Universe began and remained in the electroweak vacuum at all times, this scenario is extremely fine-tuned from the point of view of initial conditions.  This fine-tuning can be ameliorated by the exponential expansion of spacetime during inflation: intriguingly, this requires at least $\sim 40$ efolds of inflation, tantalizingly close to the $50-60$ efolds expected from horizon and flatness considerations. The Higgs could thus provide the reason for a prolonged epoch of inflation in our cosmic history.  Otherwise, the most natural initial state corresponds to our Universe initialized in the more stable but ``wrong" Higgs vacuum, and subsequently driven dynamically to the weak scale vacuum during reheating. An important, hitherto unexplored aspect of this dynamics is that the barrier between the two vacua persists when the electroweak vacuum becomes energetically favorable, becoming arbitrarily small as the temperature increases, and therefore triggers a first-order phase transition. This transition produces ultra-high (megahertz to gigahertz) frequency gravitational waves (GWs), serving as a challenging but unique SM target for GW experiments. Novel pathways for various beyond the Standard Model phenomena such as the production of dark matter and baryon asymmetry also become possible in this configuration. 
\end{abstract}

\maketitle

\section {Motivation}

According to current measurements of the top quark and Higgs boson masses, the Standard Model (SM) Higgs potential is unstable at high scales, and the electroweak (EW) vacuum that our Universe exists in is metastable. In particular, the SM Higgs potential can be written as (see e.g.\,\cite{Hook:2019zxa})
\be
V_h(h)=-\mu^2 h^2+\frac{\lambda}{4}h^4+\frac{h^6}{\Lambda_{UV}^2}\,.
\label{eq:potential}
\ee
The mass term $\mu\approx 100$ GeV is irrelevant at high scales, and will be ignored in the remainder of the paper. The quartic coupling $\lambda$, in the absence of beyond the SM (BSM) corrections, is known to run to negative values at scales higher than $\Lambda_I\gtrsim 10^{11}$ GeV, reaching $\lambda\approx -0.01$. The third (sextic) term is not present in the SM but rather represents corrections from BSM interactions of the Higgs\,\footnote{This is a generic way of capturing the effects of new physics, also used, e.g.\,in \cite{Hook:2019zxa}. Depending on the details, new physics might affect the potential in other ways, e.g. via threshold corrections to the quartic coupling, as in \cite{Espinosa:2018euj}. Such details will not qualitatively change the main results of this paper.}  at the ultraviolet (UV) scale $\Lambda_{UV}$($>\!\Lambda_I$), and is introduced to stabilize the potential, creating a stable, global (true) minimum at $v_{UV}\approx 0.04\, \Lambda_{UV}$ in addition to the electroweak (EW) minimum $v_{EW}$($\approx 0$ for the purposes of this paper). 

With two different Higgs vacua at play, the initial configuration of the Higgs field at the beginning of our Universe becomes an important question. The realization of the Higgs field in the true vacuum, or any value beyond the instability scale $\Lambda_I$, in our cosmological history is known to have potentially catastrophic consequences \cite{Ellis:2009tp,Espinosa:2007qp,Elias-Miro:2011sqh,Kobakhidze:2013tn,Hook:2014uia,Kobakhidze:2014xda,Shkerin:2015exa,Kearney:2015vba,Espinosa:2015qea,East:2016anr,Grobov:2016llk,Kohri:2016qqv,Markkanen:2018pdo,DeLuca:2022cus}. In light of this, we will henceforth refer to the deeper, true vacuum as the $\textit{wrong}$ vacuum. Such catastrophies can be avoided with nonminimal modifications of the Higgs potential (see e.g.\,\cite{Lebedev:2012sy,Fairbairn:2014zia,Herranen:2014cua,Kamada:2014ufa,Kearney:2015vba,Espinosa:2015qea,Saha:2016ozn,Enqvist:2016mqj,Ema:2017ckf,Ema:2017loe,Figueroa:2017slm,Postma:2017hbk,Fumagalli:2019ohr,Kost:2021rbi}) or reheating with very high temperatures (see e.g.\cite{Espinosa:2015qea,Kearney:2015vba,Hook:2019zxa}). In the absence of such stabilizing corrections, the safe option is to assume that the Universe began, and always remained in, the EW vacuum, which is the implicit assumption in the majority of the studies of the Higgs field in the early Universe in the literature\,\footnote{Brief forays into the region beyond the instability scale, however, can leave fascinating observable cosmological imprints \cite{Espinosa:2017sgp,Espinosa:2018euj,Espinosa:2018eve,Hook:2019zxa,Shakya:2023zvs,futurework}}. This configuration, however, is highly fine-tuned, as some generic initial configuration is far more likely to result in the Higgs settling in the wrong vacuum rather than the EW one.  For instance, if the initial Higgs field value were scanned linearly between $0$ and the Planck scale $M_P$, the probability of ending up with the Higgs at $v_\text{EW}$ is vanishingly small, $\sim \Lambda_I/M_P\approx 10^{-8}$. 

One is then left with two avenues  to naturally reconcile this conundrum with the observed Universe. The first is to begin the Universe in the EW vacuum, but invoke some means to overcome the above fine-tuning. As we will see in Section \ref{sec:inflation}, the exponential expansion of spacetime during inflation provides precisely such an opportunity, and also provides fascinating predictions for the duration and scale of inflation.
The second option is to begin the Universe in the wrong vacuum, which is the more natural initial configuration, but drive it dynamically to the EW vacuum. This can be readily achieved via reheating after high scale inflation, which gives rise to large stabilizing thermal corrections to the Higgs potential. This paper performs the first careful study of the evolution of the Higgs field from the wrong to the EW vacuum (Section \ref{sec:reheating}), pointing out, in particular, that this transition can be first-order. This first-order phase transition (FOPT) of the Higgs field features several qualitative differences compared to standard FOPTs studied in the literature, which become important for the dynamics of the new vacuum bubbles and the subsequent generation of gravitational waves; these will be addressed in Section \ref{sec:fopt}. We will briefly entertain various BSM applications of this FOPT in Section \ref{sec:bsm}, and summarize the major results of this paper in Section \ref{sec:summary}.

\section{Interplay with Inflation}
\label{sec:inflation}

We have compelling reasons to believe that our observable Universe began with a period of inflation that lasted $50-60$ e-folds in order to address the horizon and flatness problems. There are three aspects of inflation that are important for the dynamics of the Higgs field: (i) fluctuations, (ii) exponential expansion, and (iii) reheating. We will discuss the first two aspects in this section, and address the third aspect in the next section. 

\textit{Fluctuations:}
During inflation, perturbations in the metric induce fluctuations in the Higgs field of size $\partial h\sim H/(2\pi)$, where $H$ is the scale of inflation. If $H>\!\Lambda_I$, these fluctuations can drive the Higgs into the wrong vacuum even if it begins in the EW vacuum \cite{Ellis:2009tp,Espinosa:2007qp,Elias-Miro:2011sqh,Kobakhidze:2013tn,Hook:2014uia,Kobakhidze:2014xda,Shkerin:2015exa,Kearney:2015vba,Espinosa:2015qea,East:2016anr,Grobov:2016llk,Kohri:2016qqv,Markkanen:2018pdo,DeLuca:2022cus}. A Fokker-Planck approach reveals that the wrong vacuum becomes exponentially favored in such cases \cite{Hook:2014uia}, thereby worsening the fine-tuning issue in high-scale inflation scenarios. 

\textit{Exponential Expansion:} 
During inflation, spacetime expands exponentially, with the rate of expansion determined by the Hubble scale in the local (causally connected) patch 
\be
H=\left(\frac{8\pi(V_\phi+V_h(h))}{3  M_P^2}\right)^{1/2},
\ee
where $V_\phi$ is the inflaton energy density\,\footnote{For some orthogonal studies where a metastable, positive energy higher Higgs vacuum provides the setting for inflation, see \cite{Masina:2011un,Masina:2011aa,Masina:2012yd}.}. Note that the Higgs potential energy also contributes, with $V_h(v_{EW})\approx 0$ and $V_h(v_{UV})\approx -0.001 v_{UV}^4<0$. Indeed, if $V_\phi+V_h(v_{UV})< 0$, the regions of spacetime with the Higgs in the wrong vacuum fail to inflate, and instead become regions of anti de-Sitter space (AdS) that simply crunch away, leaving the EW vacuum regions as the only surviving spacetimes\,\footnote{This idea has been used in various papers to rescue otherwise-fine-tuned vacuum configurations \cite{Hook:2014uia, Espinosa:2015qea,Cheung:2018xnu,Csaki:2020zqz}.}. However, these AdS regions of the wrong Higgs vacuum are not completely benign: they tend to draw the ``good" (EW) vacuum regions into the AdS regime even as they collapse \cite{East:2016anr}. Nevertheless, as long as no wrong vacuum region existed in our past light-cone, our current Universe remains safe from such catastrophies. 

The more interesting configuration occurs if $V_\phi+V_h(v_{UV})> 0$. In this case, both the EW and wrong vacua inflate\,\footnote{We have assumed that all regions with the initial Higgs configuration $h>\Lambda_I$ classically roll to the wrong vacuum. This rolling, especially from very large field values, might induce large perturbations that change the equation of state and prevent inflation in these regions; we ignore this effect here for simplicity.}, but the former inflates faster because of higher potential energy, and hence grows in relative volume measure by a factor $e^{(H_\text{EW}-H_\text{UV}) t}$ as inflation proceeds (where $H_\text{EW(UV)}$ corresponds to the scale of inflation in the spacetime regions where the Higgs field sits in the EW (UV) vacuum). Recall that from the point of view of initial conditions, the probability of finding a region with the Higgs in the EW vacuum is minuscule, $\sim \Lambda_I/M_P\approx 10^{-8}$. A period of unequal expansion during inflation can thus overcome this suppression. One can calculate that the number of e-folds of inflation needed for the volume of spacetime with the Higgs in the EW vacuum to be larger than the volume in the wrong vacuum is (after a bit of algebra) 
\be
n_{\text{e-folds}}> \frac{2 V_\phi}{|V_h(h_t)|}\text{ln}\left(\frac{M_P}{\Lambda_I}\right)\,.
\ee
If $V_\phi\gtrsim V_h(v_\text{UV})$ and $M_P/\Lambda_I\approx 10^8$, the above implies $n_{\text{efolds}}\!\gtrsim\!37$. This is intriguingly close to the $50-60$ efolds of inflation needed to address the infamous horizon and flatness problems. The Higgs metastability could therefore, incredibly, play a key role in explaining these fundamental properties of our Universe!

Note that the above is simply the most optimistic lower limit; if $V_\phi\gg V_h(v_\text{UV})$, then obviously $n_{\text{efolds}} \gg 37$ is required as the difference between the potential energies (and thus Hubble rates) in spacetime regions featuring different Higgs minima becomes negligibly small, requiring a longer period of inflation to make the EW vacuum more voluminous. If inflation does not last so long, then the Higgs sits in the wrong vacuum in most regions, and the fine-tuning issue persists. A large reheating effect is then required to push the Higgs to the EW vacuum. This process will be studied in the next section.

\section{Dynamics during reheating}
\label{sec:reheating}

When inflation ends, the inflaton energy density gets converted into a radiation bath, creating thermal corrections to the Higgs potential. If the temperature is sufficiently high, these corrections lift up the wrong Higgs vacuum, causing the Higgs field to classically roll to the EW minimum (see e.g.\cite{Espinosa:2015qea,Kearney:2015vba,Hook:2019zxa}). However, these studies simply assumed that the thermal corrections turn on very rapidly, essentially instantaneously, without paying careful attention to the details of the reheating process. In reality, the temperature increases gradually, dictated by the decay width of the inflaton (or the reheaton, if reheating into the SM sector is controlled by a field other than the inflaton). As we will see in this section, such details can in fact carry extremely important implications for how our cosmological history unfolds.

The thermal correction to the Higgs potential can be parameterized as \cite{Espinosa:2017sgp,Franciolini:2018ebs,Hook:2019zxa}
\be
\Delta V_T(h,T)\approx0.06 \,h^2\, T^2\, e^{-h/(2\pi T)}\,.
\label{eq:thcorr}
\ee
This correction is only valid for $h\!<\!2\pi T$, and the exponential term serves the practical purpose of shutting it off rapidly at larger field values (and is known to be accurate for $h < 10\, T$ \cite{Espinosa:2017sgp}).\,\footnote{This formulation implicitly assumes that the particles in the radiation bath are in thermal equilibrium, which might not be satisfied at early times. The scattering rate for the thermalization of SM particles through weak interactions is $\Gamma_{th}\sim g^2 T$, which can be smaller than the inverse timescale over which the thermal corrections are relevant when $T$ is small. Furthermore, when the Higgs is sitting at the wrong minimum with a large vev, the SM particles are correspondingly heavier, with masses possibly larger than $T$, further preventing full thermalization. Here we ignore such caveats and assume that the thermal correction is reasonably captured by Eq.\,\ref{eq:thcorr}.} 

\begin{figure}[t]
\begin{center}
 \includegraphics[width=0.99\linewidth]{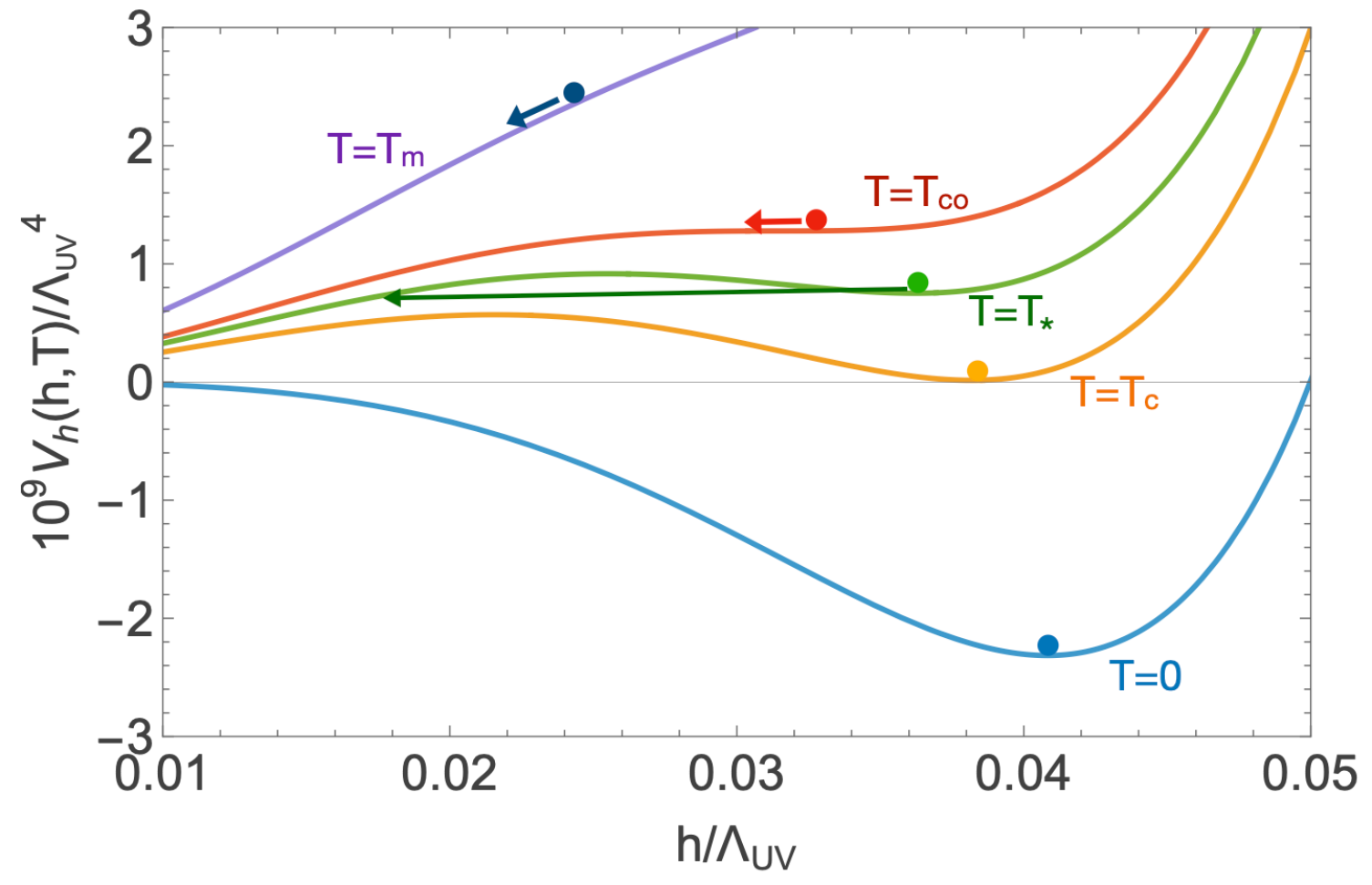}
  \end{center}\vspace{-4mm}
\caption{Thermally corrected Higgs potential for increasing temperatures: zero temperature (blue); critical temperature $T_c$ where the two vacua become degenerate (orange); nucleation temperature $T_*$ where the phase transition becomes efficient (green), crossover temperature $T_{co}$ where the higher minimum disappears (red), and maximum temperature $T_m$ reached by the radiation bath (purple).}
\label{fig:potentials}
\end{figure}

Fig.\,\ref{fig:potentials} shows various crucial intermediate stages of the thermally corrected Higgs potential $V_h(h,T)=V_h(h)+\Delta V_T(h,T)$ as the temperature increases. At zero temperature (blue curve), the Higgs field sits at the wrong minimum at $h\approx 0.04 \Lambda_{UV}$. As the temperature is turned on, the thermal correction lifts this stable minimum relative to the EW minimum (which is at $V_h(v_{EW})\approx 0$, not visible in the figure). At the critical temperature $T_c$ (orange curve), the two vacua become degenerate in energy. An important point worth highlighting here is that \textit{the barrier between the two vacua persists at $T\gtrsim T_c$} thanks to the interplay between the positive thermal correction and the negative Higgs quartic term. This opens the possibility of a \textit{first-order phase transition} (FOPT), where the Higgs field tunnels through the classically impenetrable barrier to the EW vacuum. However, the temperature needs to rise further to the nucleation temperature $T_*$ (green curve) for the FOPT to be efficient (this needs to be calculated more carefully than in standard FOPT scenarios, as will be discussed below). If the temperature rises further to the crossover temperature $T_{co}$ (red curve), the higher minimum disappears completely; beyond this, we get a crossover transition (where the Higgs field simply rolls classically down to the EW vacuum) rather than a FOPT. This would be the case at the highest temperature $T_m$ (purple curve) if $T_m\!>\!T_{co}$. Earlier studies of reheating in the context of the Higgs metastability assumed an instantaneous jump from $T=0$ to $T_m\!>\!T_{co}$ and therefore completely missed the possibility of a FOPT at intermediate temperatures.

Next, we derive the conditions for the FOPT to be realized in this setup. Note that simply obtaining a configuration such as the green curve in Fig.\,\ref{fig:potentials} is not sufficient;  the tunneling rate might be too small, and the Higgs potential might evolve to a configuration that prohibits a FOPT (such as the red curve) too rapidly to make a FOPT viable. 

The probability of nucleating a critical bubble of the more stable vacuum in a background of the metastable vacuum per unit volume at finite temperature is given by
\be
\Gamma_\text{nuc}\approx T^4 \left(\frac{S_3}{2\pi T}\right)^{3/2} e^{-S_3/T}\,,
\label{eq:condition}
\ee
where $S_3/T$ is the $\mathcal{O}(3)$-symmetric bounce action. There exist approximate analytic formulae for $S_3/T$ in the thin-wall limit (see e.g.\cite{Buen-Abad:2023hex}); however, the thin-wall approximation is not applicable to the SM Higgs potential under consideration, hence $S_3/T$ must be evaluated numerically. We calculate $S_3/T$ using the numerical package {\tt{FindBounce}} \cite{Guada:2020xnz,Guada:2018jek}. 

The FOPT is generally understood to become viable when $\Gamma_\text{nuc}> H^4$, i.e.\,more than one critical bubble nucleates per Hubble volume per Hubble time. This condition, while satisfactory in standard FOPT calculations in a radiation dominated or supercooled setup, might be insufficient when the temperature is increasing: in such cases, the temperature might increase from $T_*$ to $T_{co}$ (see Fig.\,\ref{fig:potentials}) in a shorter timescale $\Delta t<H^{-1}$, hence a critical bubble needs to nucleate per Hubble volume within $\Delta t$ rather than a Hubble time, otherwise in most regions of spacetime the phase transition would occur via the Higgs rolling classically to the EW vacuum configuration (which becomes possible at $T>T_{co}$) instead of bubble nucleation. We thus need to enforce a stronger condition for a successful FOPT
\be
\Gamma_\text{nuc}> H^3/\Delta t,
\label{eq:newcondition}
\ee
where $\Delta t$ is the time taken for the temperature to change from $T_*$ to $T_{co}$, and $T_*$ is the lowest temperature at which the above relation is first satisfied. 

\begin{figure}[t]
\begin{center}
 \includegraphics[width=.99\linewidth]{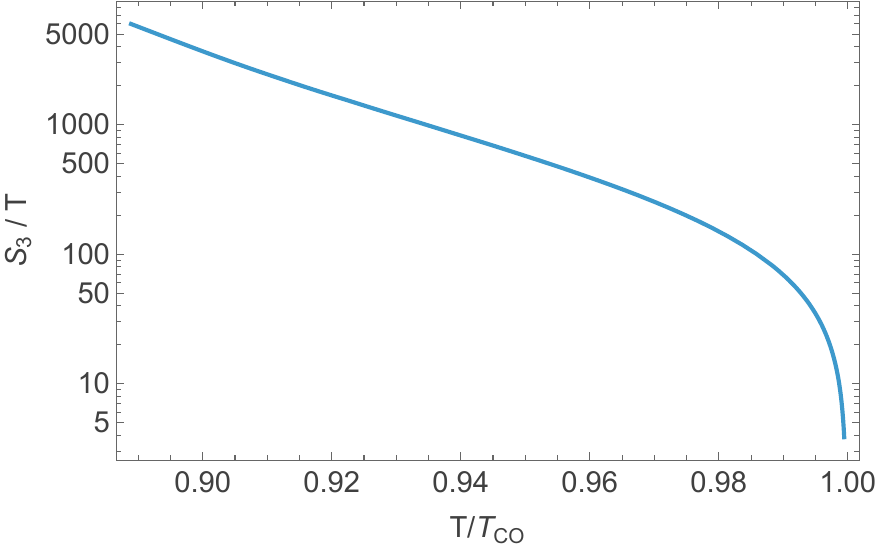}
  \end{center}\vspace{-4mm}
\caption{The bounce action $S_3/T$ as a function of temperature. As $T\to T_{co}$, the potential barrier separating the two vacua becomes vanishingly small, quickly driving the action to $\mathcal{O}(1)$ values, inevitably leading to rapid tunneling.}
\label{fig:action}
\end{figure}

Here one might worry that Eq.\,\ref{eq:newcondition} is too stringent to be satisfied if $\Delta t H\!\ll\!1$. However, one can make the reassuring observation that as $T$ approaches $T_{co}$, the potential barrier becomes vanishingly small and $S_3/T$ inevitably approaches $\mathcal{O}(1)$ values; this can be seen clearly in Fig.\,\ref{fig:action}. We can then reasonably expect Eq.\,\ref{eq:newcondition} to hold at some point shortly before  $T$ reaches $T_{co}$, triggering the FOPT successfully. To be more specific, when $S_3/T$ becomes very small as $T\to T_{co}$, we can write Eq.\,\ref{eq:condition} as $\Gamma_\text{nuc}\approx T^4$, and Eq.\,\ref{eq:newcondition} can subsequently be rewritten as 
\be
\Delta t\,H > (H/T)^4 > g_*^2 (T/M_P)^4\,,
\ee
where $g_*$ is the number of relativistic degrees of freedom in the bath. When $T$ is many orders of magnitude below the Planck scale, this condition can, in general, be comfortably satisfied. 

This point deserves special emphasis: the inevitable vanishing of the potential barrier as the temperature increases to the crossover temperature is a unique feature of the reheating setup not encountered in the standard (cooling) FOPT setups, where a sizable potential barrier is generally present when the true vacuum becomes energetically favorable. As a consequence, while a FOPT that can complete is not guaranteed in cooling setups, it is almost always guaranteed in a setting with increasing temperature (provided the temperature reaches the crossover temperature). 

Note that we simply need at least one critical bubble nucleated per Hubble volume in the time $\Delta t$, but the phase transition does not need to complete within this timeframe: even if the temperature rises beyond $T_{co}$ in the region between the bubbles and the Higgs begins to roll classically towards the EW vacuum here, the classical evolution is uneven across the bubble walls, hence the Higgs gradient at the bubble wall does not get erased (until the Higgs field reaches the EW minimum on both sides of the wall). The bubble walls thus persist until they collide with each other. Therefore, the duration of the phase transition $\beta^{-1}$ could in principle be longer than $\Delta t$. 

$\beta$ can be calculated using the standard formula
\be
\beta=\frac{dS_3}{dt}=\frac{dS_3}{dT}\frac{dT}{dt}\,.
\ee
From this, we see that the time-temperature relation is crucial in determining the viability as well as properties of the FOPT. This depends on the reheating process: if the inflaton decays with a decay rate $\Gamma_\phi$ into the SM thermal bath, then the energy densities in the inflaton and SM radiation ($\rho_\phi$ and $\rho_r$, respectively) are given by the following set of coupled differential equations
\bea
\dot{\rho}_\phi+3H\rho_\phi &=&-\Gamma_\phi \,\rho_\phi, \nonumber\\
\dot{\rho}_r +4H\rho_r &=& \Gamma_\phi \,\rho_\phi\,.
\eea
Note that this takes into account redshifting due to Hubble expansion\,\footnote{These formulae assume that the radiation bath consists of relativistic particles. This is a reasonable approximation as the photon remains massless even in the wrong vacuum; furthermore, if reheating completes within $\mathcal{O}$(Hubble time), then most of the SM particles (other than the top quark and the $W, Z, h$ bosons) also become relativistic.}. The temperature of the SM bath is then determined by $\rho_r=\frac{\pi^2}{30}g_* T^4$. 

\begin{figure}[t]
\begin{center}
 \includegraphics[width=.99\linewidth]{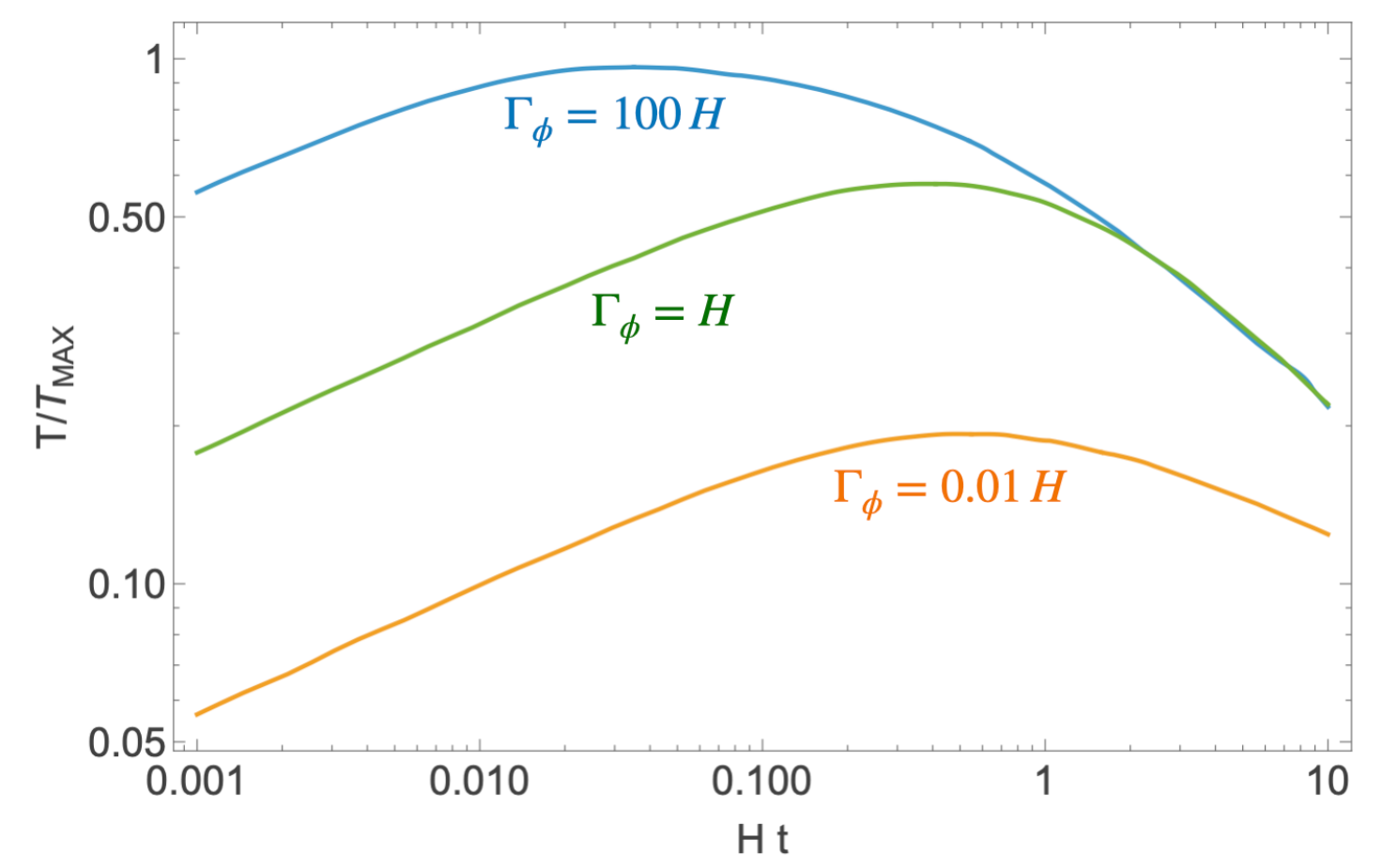}
  \end{center}\vspace{-4mm}
\caption{Temperature of radiation bath as a function of time for various choices of inflaton decay rate.}
\label{fig:temperature}
\end{figure}

Fig.\,\ref{fig:temperature} shows the temperature-time relation for various choices of $\Gamma_\phi$.  Here $T_\text{MAX}$ is the temperature reached if reheating occurs instantaneously.  As expected, $T$ reaches $T_{RH}$ if the inflaton decays within a Hubble time, otherwise redshifting due to Hubble expansion always causes the temperature to drop beyond $H t\!>\!1$. We therefore expect the FOPT to occur on the rising portion of these curves, within a Hubble time.

There are two qualitatively different cases to consider:
\begin{itemize}
\item \textit{Case I}: FOPT occurs on the linearly rising part of the $T-t$ curve. Here we can write $\rho_r\approx \Gamma_\phi\rho_{\phi,i} t$, where $\rho_{\phi,i}$ is the initial inflaton density, as the inflaton energy density is approximately constant in this regime. This enables us to calculate $dT/dt$ analytically in terms of the Hubble scale, eventually obtaining a fairly simple relation $\beta/H\approx T \frac{dS_3}{dT}$. 
\item \textit{Case II}: FOPT occurs close to the top of the $T-t$ curve. In this case $dT/dt$ and $\beta$ need to be evaluated numerically. Note that $dT/dt$ here is smaller than in Case I: the temperature, hence also the action, change more slowly, and the phase transition completes over a longer time period. 
\end{itemize}

For Case I, numerical evaluation yields $\beta/H\sim 7000$ and $T_*\approx T_{co}\approx 0.009\, \Lambda_{UV}$ for $\Gamma_\phi=H$, with the now metastable high scale minimum at $h_*\approx 0.033 \Lambda_{UV}$. These numbers are found to be largely independent of $\Lambda_{UV}$. Note that $\beta/H\gtrsim \mathcal{O}(1000)$ is typical for thermal FOPTs, where the action changes rapidly as the temperature changes, causing the FOPT to complete in far less than a Hubble time \cite{Breitbach:2018ddu}. 

For Case II, it should be possible to get $\beta/H=\mathcal{O}(10-100)$ with an appropriate choice of parameters. However, it is worth noting that Case II is a somewhat fine-tuned scenario as it requires $T_*\approx T_m$, i.e. the phase transition is triggered precisely as the temperature of the radiation bath reaches its maximum value. In this regime the Universe also faces another danger: the thermal correction might get suppressed before the Higgs field has had a chance to roll past the instability scale towards the EW minimum. In this case, it will classically roll back to the wrong vacuum: in physical space, the nucleated bubbles would collapse instead of expanding, and the transition fails to complete. 

In both cases, one might also worry that the Higgs might tunnel back to the wrong minimum when the temperature inevitably drops and the wrong vacuum becomes energetically favored again. However, when this happens, the barrier between the two minima is significant, and the tunneling probability is severely suppressed, as can be checked explicitly by numerically calculating the bounce action.

\section {Properties of the First-Order Phase Transition}
\label{sec:fopt}

Having established that the SM Higgs field can undergo a FOPT from the wrong vacuum to the EW vacuum, in this section we will discuss various properties of the FOPT, in particular the stochastic gravitational wave (GW) signal that is generated.\,\footnote{For a recent study of FOPTs during reheating, see \cite{Buen-Abad:2023hex}.} 

\begin{figure}[t]
\begin{center}
 \includegraphics[width=0.99\linewidth]{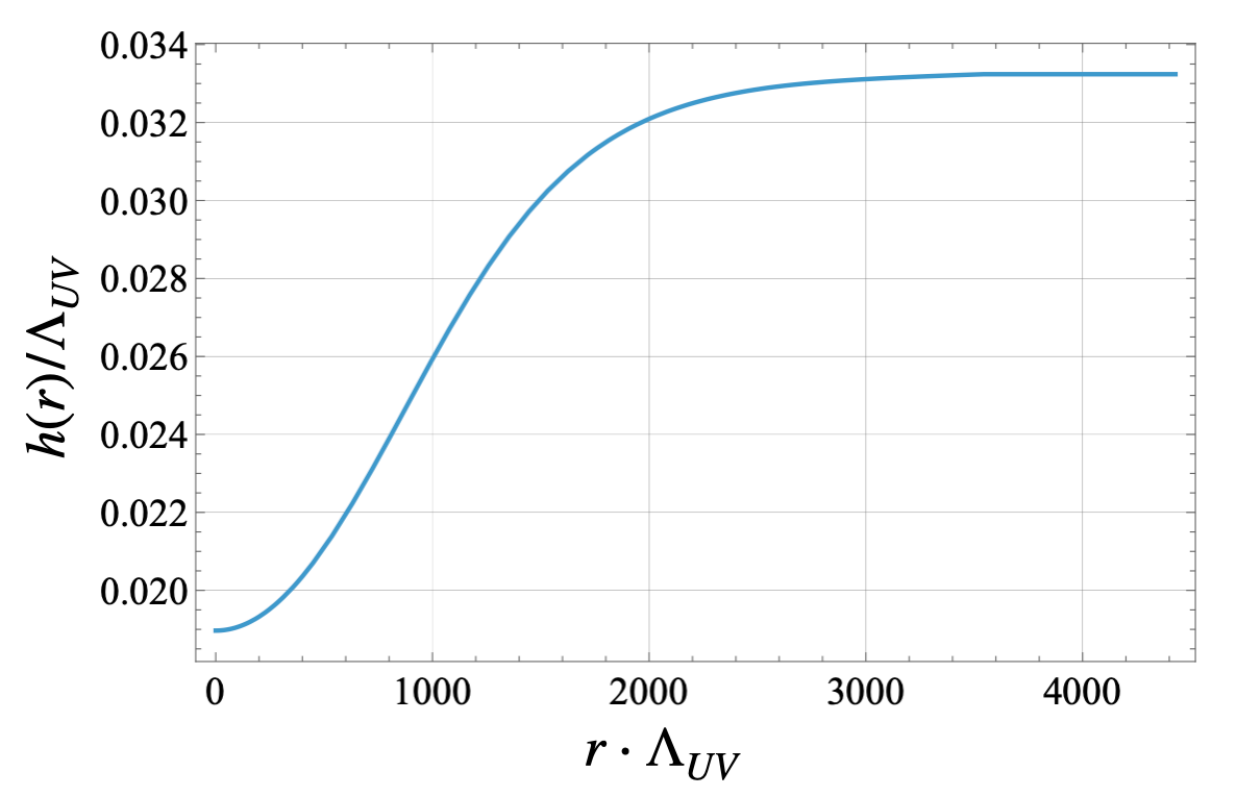}
  \end{center}\vspace{-4mm}
\caption{Higgs field profile in a critical bubble.}
\label{fig:profile}
\end{figure}

We begin by plotting the Higgs field profile in a critical bubble in Fig.\,\ref{fig:profile}, obtained using the numerical package {\tt{FindBounce}}. This shows that the field tunnels to a point $h_c\approx 0.02 \Lambda_{UV}$ rather than the EW minimum; in other words, this is a thick wall bubble. As the bubble expands, the interior of the bubble wall will feature Higgs field oscillations as the field rolls classically from $h_c$ to the EW minimum, carrying a significant fraction of the energy released in the transition. The bubble radius at nucleation is seen to be $\approx 3000 /\Lambda_{UV}\approx 30/T_*$.

The fraction of energy carried by the bubble walls (relevant for the subsequent generation of GWs) can be parametrized as 
\be
\alpha=\frac{V_h(h_*,T_*)-V_h(h_c,T_*)}{\rho_r(\text{max})}\,,
\ee
where $\rho_r(\text{max})$ corresponds to the maximum energy density in radiation. Recall that there is a lower bound $\rho_r(\text{max})> \frac{\pi^2}{30}g_* T_*^4$ in order for the FOPT to complete. Meanwhile, the numerator numerically evaluates to $\approx 3\times 10^{-10}\Lambda_{UV}^4$. Combining this with the above lower bound on $\rho_r(\text{max})$ implies $\alpha<10^{-3}$. 

To deduce various phenomenological features of the FOPT, we also need to determine the bubble wall velocity. This is particularly crucial for deducing the GW signal produced by the FOPT. If the bubble walls run away, the GWs are primarily sourced by the scalar field energy densities during bubble collision\,\cite{Kosowsky:1991ua,Kosowsky:1992rz,Kosowsky:1992vn,Kamionkowski:1993fg,Caprini:2007xq,Huber:2008hg,Bodeker:2009qy,Jinno:2016vai,Jinno:2017fby,Konstandin:2017sat,Cutting:2018tjt,Cutting:2020nla,Lewicki:2020azd} or by the distribution of particles produced from bubble collisions \cite{Inomata:2024rkt}. On the other hand, if the bubble walls reach terminal velocity due to significant friction from the plasma surrounding the bubbles, GWs instead arise from sound waves\,\cite{Hindmarsh:2013xza,Hindmarsh:2015qta,Hindmarsh:2017gnf,Cutting:2019zws,Hindmarsh:2016lnk,Hindmarsh:2019phv} and turbulence\,\cite{Kamionkowski:1993fg,Caprini:2009yp,Brandenburg:2017neh,Cutting:2019zws,RoperPol:2019wvy,Dahl:2021wyk,Auclair:2022jod} in the plasma, or via distributions of particles that do not thermalize\,\cite{Jinno:2022fom}.

In standard (symmetry-breaking) phase transitions involving sectors with gauge bosons, bubble walls reach a terminal velocity due to transition radiation, which produces a pressure on bubble walls that grows linearly with the Lorentz boost factor $\gamma_w$ of the wall \cite{Bodeker:2017cim,Hoche:2020ysm,Gouttenoire:2021kjv,Ai:2024shx,Long:2024sqg,Ai:2025bjw}. The FOPT in question here is different: it is a symmetry-\textit{restoring} phase transition, where the particles outside the bubbles are massive and lose mass as they enter the bubbles. Transition radiation in such configurations has been studied in \cite{frictionsymmres}, and found to produce $\textit{negative}$ pressure at low $\gamma_w$, only recovering the more familiar positive scaling proportional to $\gamma_w$ at large $\gamma_w\gtrsim \mathcal{O}(10)$ (see also \cite{Azatov:2024auq}). The Higgs bubbles of EW vacuum are therefore expected to reach terminal velocities with  $\gamma_w\approx \mathcal{O}(10)$ despite significant interactions between the Higgs bubble walls and the surrounding SM plasma.

Consequently, GWs from this FOPT are primarily produced by sound waves\,\footnote{A secondary GW contribution arises from the oscillations of the Higgs field inside the bubbles \cite{Cutting:2018tjt,Cutting:2020nla}, but this will be several orders of magnitude higher in frequency, and more challenging to detect.}. For $\beta/H=7000$ as found above, and $\Lambda_{UV}=10^{14}$ GeV, the GW signal has a peak frequency today at $\sim10$ GHz. Varying $\Lambda_{UV}$ between $10^{12}$ to $10^{15}$ GeV, and allowing for $\beta/H$ between $\mathcal{O}(10)$ (as can occur in Case II above) and $\mathcal{O}(10^4)$, the present day peak frequency spans the range MHz$-100$ GHz. Likewise, the present day peak amplitude of the signal is approximately
$\Omega_\text{GW}\approx 10^{-6} \alpha^2 (H/\beta)$ \cite{Caprini:2024hue}. Since $\alpha<10^{-3}$, this implies $\Omega_\text{GW}<10^{-15}$ for Case I (with $\beta/H\approx 7000$) and $\Omega_\text{GW}<10^{-13}$ for Case I (with $\beta/H\approx 10$). Such high frequencies and low amplitudes imply that detecting such GW signals with future experimental efforts will be extremely challenging; nevertheless, this serves as a unique and tantalizing \textit{Standard Model} benchmark target for experiments to aim for. 


\section{Implications for BSM Physics}
\label{sec:bsm}

The first obvious phenomenological question, given the realization of a FOPT with the SM Higgs, is whether this configuration can produce the baryon asymmetry of the Universe via electroweak baryogenesis. This is unfortunately not possible due to two primary reasons: (i) Electroweak baryogenesis requires slow-moving bubble walls ($v\sim 0.1$) in order to efficiently produce and preserve the asymmetry; in our case, we have the walls moving relativistically with $v\approx 1$; (ii) since this FOPT occurs at very high temperatures $T>\Lambda_I\sim 10^{11}$ GeV, the electroweak sphalerons remain active after the transition, and would wash out any asymmetry that was produced in the manner of electroweak baryogenesis. 

Nevertheless, the realization of ultra-relativistic bubble walls might provide new pathways to producing the baryon asymmetry as well as other BSM applications, such as the production of dark matter, as explored in various generic setups \cite{Azatov:2021ifm,Baldes:2021vyz,Azatov:2021irb,Baldes:2022oev,Chun:2023ezg,Mansour:2023fwj,Shakya:2023kjf,Ai:2024ikj,Giudice:2024tcp,Cataldi:2024pgt}. This is because bubble collisions and bubble-plasma interactions have access to a higher energy scale $\gamma_w T\sim \mathcal{O}(10) T$ than the temperature of the ambient plasma.  Such ideas in the context of the SM Higgs high scale FOPT will be explored in detail in a companion paper \cite{futurework}.

\section{Summary}
\label{sec:summary}

This section summarizes the main ideas presented in this paper. 

\begin{figure}[t]
\begin{center}
 \includegraphics[width=0.9\linewidth]{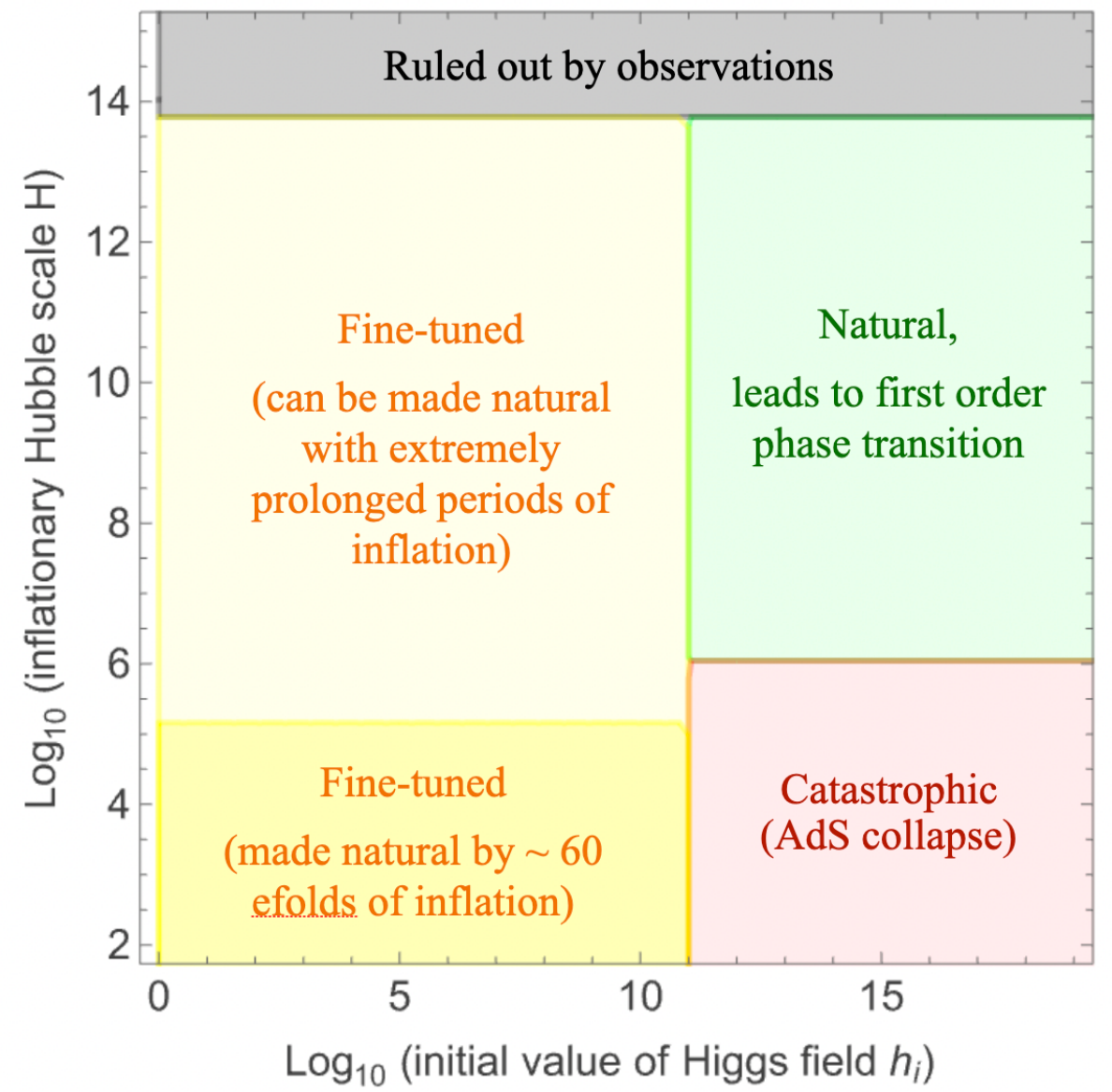}
  \end{center}\vspace{-4mm}
\caption{Various possibilities for the initial configuration of the Higgs field ($h_i$) at the beginning of the Universe and its interplay with the scale of inflation $H$, for $\Lambda_I=10^{11}$ GeV and $\Lambda_{UV}=10^{14}$ GeV.}
\label{fig:paramspace}
\end{figure}

The Standard Model Higgs potential features a deeper (``wrong") minimum beyond the instability scale $\Lambda_I$ at large field value $v_{UV}\approx0.04\Lambda_{UV}$, which gives rise to three qualitatively different options for the initial configuration of the Higgs field $h_i$ and its interplay with inflation at scale $H$ (as shown schematically in Fig.\,\ref{fig:paramspace} for $\Lambda_I=10^{11}$ GeV and $\Lambda_{UV}=10^{14}$ GeV):
\begin{itemize}
\item The catastrophic configuration $h_i\!>\!\lambda_I$ and $H< {v_{UV}^2/M_P}$ (red region): the Universe sits in the wrong minimum but reheating effects are not strong enough to lift the Higgs to the EW vacuum, making the existence of a Universe such as ours impossible. 
\item The fine-tuned configuration $h_i\!<\!\Lambda_I$ (yellow region): The Higgs field can survive in the EW vacuum, but this initial condition is fine-tuned to 1 part in $10^8$ if the initial value were randomly picked from a linear distribution. The fine-tuning can be overcome if the wrong vacuum patches collapse due to negative energy density, or if a period of inflation preferentially inflates the EW vacuum regions over the wrong vacuum regions (Sec.\,\ref{sec:inflation}). In the figure, we distinguish the region where $\sim 60$ e-folds of inflation would be sufficient to overcome the initial fine-tuning (lower, darker yellow) from the part of parameter space where substantially longer periods of inflation would be needed (upper, lighter yellow). The Higgs metastability could thus provide the reason for a prolonged epoch of inflation in our cosmic history. 
\item The natural configuration $h_i>\Lambda_I,~H> {v_{UV}^2/M_P}$: This is the more natural initial configuration for the Higgs field, corresponding to it sitting in the wrong minimum. A sufficiently large reheat temperature post-inflation creates the required thermal corrections to lift the wrong vacuum and drive the Higgs field dynamically to the EW minimum. We have seen that this is realized via a first-order phase transition (FOPT) (Section \ref{sec:reheating}).  
\end{itemize}

The FOPT has several salient properties:
\begin{itemize}
\item As the temperature rises, the potential barrier between the two vacua in the thermally corrected Higgs potential can be made arbitrarily small, significantly enhancing the tunneling rate. A FOPT is therefore realized no matter how fast reheating occurs. Note that this is a feature unique to transitions with increasing temperature; moreover, it is applicable to $\textit{any}$ scalar potential during reheating, not just the SM Higgs field. Requiring a successful FOPT also enforces a lower bound on the reheat temperature $T_{RH}> 0.009\, \Lambda_{UV}$. 
\item The FOPT produces gravitational waves signals with present day peak frequency in the MHz$-100$\,GHz range and peak amplitude $\Omega_{GW}\!<\!10^{-13}$. This provides an extremely challenging but a very well-motivated and compelling SM target for future ultra-high frequency GW experimental efforts. 
\item Despite significant interactions between the thermal plasma and the bubbles, the bubble walls attain partial runaway behavior due to the modified nature of plasma friction in this symmetry-restoring transition, reaching ultrarelativistic terminal velocities with Lorentz boost factor $\gamma_w\approx \mathcal{O}(10)$. While incompatible with electroweak baryogenesis, this could open up novel BSM applications. 
\end{itemize}

Many of these intriguing aspects will be explored in greater detail in a series of upcoming papers \cite{futurework}.

\section*{Acknowledgments} 

The author acknowledges helpful conversations and/or feedback on the manuscript from Gian Giudice, Anson Hook, Marc Kamionkowski, Davide Racco, Geraldine Servant, and James Wells. This work is supported by the Deutsche Forschungsgemeinschaft under Germany's Excellence Strategy - EXC 2121 Quantum Universe - 390833306. 

\bibliography{higgsuniverse}{}

\end{document}